# HOPE: Holistic STT-RAM Architecture Exploration Framework for Future Cross-Platform Analysis

**SAEED SEYEDFARAJI (Graduate Student Member, IEEE), MARKUS BICHL, ASAD AFTAB (Graduate Student Member, IEEE),and SEMEEN REHMAN(Member, IEEE).[1],**

[1]Faculty of Electrical Engineering and Information Technology, Vienna University of Technology (TU-Wien), 1040 Vienna, Austria

Corresponding author: Saeed Seyedfaraji (e-mail: saeed.seyedfaraji@tuwien.ac.at).

**ABSTRACT** Spin Transfer Torque Random Access Memory (STT-RAM) is an emerging Non-Volatile Memory (NVM) technology that has garnered attention to overcome the drawbacks of conventional CMOS-based technologies. However, such technologies must be evaluated before deployment under real workloads and architecture. But there is a lack of available open-source STT-RAM-based system evaluation framework, which hampers research and experimentation and impacts the adoption of STT-RAM in a system. This paper proposes a novel, extendable STT-RAM memory controller design integrated inside the gem5 simulator. Our framework enables understanding various aspects of STT-RAM, i.e., power, delay, clock cycles, energy, and system throughput. We will open-source our HOPE framework, which will fuel research and aid in accelerating the development of future system architectures based on STT-RAM. It will also facilitate the user for further tool enhancement.

## I. INTRODUCTION

STT-RAM boasts several compelling features including non-volatility, high density, soft error reliability, CMOS compatibility, high endurance, and scalability [1]–[4]. According to the International Roadmap for Devices and Systems (IRDS) [5], STT-RAM emerges as the most promising memory option to replace conventional memory technologies. Table 1 presents a comprehensive comparison of various design elements related to memory technologies, such as endurance, associated read/write energy, latency, and compatibility with CMOS technology size. It is worth noting that the listed suppliers are not exclusive options, as alternative providers may also offer each type of memory technology. Considering the decreasing technology node size over time due to Denard's law, it is important to note that the suggested compatibility in the table is based on the findings from the literature review of [1]–[4], [6]–[18]. Therefore, it is plausible that each memory technology could potentially be fabricated with other CMOS sizes.

Comparatively, the write and read energy of Dynamic Random Access Memory (DRAM) Static Random Access Memory (SRAM) per operation is lower than that of NVMs. Nevertheless, both DRAM and SRAM, as volatile memory types, rely on a continuous power supply to retain stored data. DRAM, commonly employed as the main memory in computers, consumes more energy due to its constant need for refreshing to preserve data integrity. Similarly, SRAM, another volatile memory technology, also requires a consistent power supply for data retention. On the other hand, NVMs like Flash memory can maintain data even when power is off, resulting in lower energy consumption compared to DRAM and SRAM. Therefore, NVM became an attractive alternative for main memories because of lower energy consumption. Phase Change Memory (PCM), and STT-RAM [19] are some of the common NVMs proposed to replace DRAM as main memory. They have also been explored for building larger on-chip caches because of their high density. Amongst all, STT-RAM has matured a lot in terms of its on-chip computation, and high energy efficiency [20].

However, the utilization of STT-RAM in widespread industrial applications is hindered by several limitations. Notably, challenges such as write operation delays and high write energy consumption. To overcome these issues, different approaches have been explored at various levels of abstraction, including circuit-level approaches [21]–[25], architecture-level approaches [10], [26], [27], and methods proposed from an application perspective [28], [29].

TABLE 1: Comprehensive comparison of NVM technologies [1]–[4], [6]–[18]

| | STT-RAM | PCMRAM | RRAM | Fe-FET | FLASH | SRAM | DRAM |
|---|---|---|---|---|---|---|---|
| Non-volatility | + | + | + | + | + | - | - |
| Data Retention (years) | 10 | 10 | 10 | 10 | 10 | - | - |
| Cell Endurance (cycles) | $10^{16}$ | $10^{12}$ | $10^{10}$ | $10^{12}$ | $10^{6}$ | $10^{16}$ | $10^{15}$ |
| Cell Size ($F^2$) | 6-20 | 4-8 | 4 | 4-8 | 4-6 | >100 | 4-12 |
| Technology node (nm) | 45 | 65 | 40 | 5 | 15 | 10 | 32 |
| Read Latency (ns) | 2-20 | 20-50 | <10 | 10 | $25 \times 10^{3}$ | <5 | 2 |
| Write Latency (ns) | 5-20 | 30 | 5 | 10 | $500 \times 10^{3}$ | <5 | 5 |
| Erase Latency (ns) | 5-20 | 30 | 10 | 10 | 2 (ms) | <5 | 5 |
| Write Energy (pJ) | 0.1-2.5 | 18 | 0.1 | 1 | 0.1 - 1 | <0.1 | <0.1 |
| Erase Energy (pJ) | 1 | 18 | 0.1 | 1 | 1000 | <1 | <1 |
| Suppliers | Toshiba, Hitachi | Samsung, Intel, WD, IBM | Panasonic, Micron | Globalfoundries, FMC | Micron, Samsung | Qualcomm, Intel | Samsung, SK Hynix |

### A. NEED FOR STT-RAM BASED SYSTEM EVALUATION FRAMEWORKS

Researchers have made significant contributions to enhancing comparison metrics within their respective levels of abstraction [1], [2], [4], [6]–[11], [13], [14], [24], [25]. However, it is crucial to note that the current architectural perspective findings are derived from a behavioral model of the circuit, which may not offer precise and detailed outcomes comparable to those from a real computing system. This approach falls short in addressing the need for a comprehensive system exploration framework. Hence, the associated research challenge is *how to design a holistic system evaluation framework that can be used to evaluate the impact of incorporating STT-RAM memories in current systems, while accurately modeling the scaling, energy consumption and performance characteristics of these devices and enabling architectural design space exploration.*

A number of simulation environments are available for research and development of system-level exploration of computer architectures, such as gem5 and ZSIM [30]. However, gem5 is widely used due to its ability to emulate the full-system mode and help in the exploration of system-level metrics, with different instruction set architectures (ISAs) such as Alpha, ARM, SPARC, MIPS, RISC-V, and x86 ISAs), and various timing and CPU modes [31], [32]. ZSIM, as an alternative simulation software, does not offer full-system simulation capabilities, but also does not rely on event-driven execution and is therefore faster. As this work targets the integration of STT-RAM into a complete system, also showing capabilities of executing an operating system on top of STT-RAM, gem5 is the selected choice. gem5 showed fast enough simulation speed for benchmark applications on top of an operating system.

### B. ANALYZING STT-RAM IMPACT ON DIFFERENT APPLICATIONS

In order to verify the framework design along with its advantages, a case study i.e., investigating STT-RAM from the perspective of reducing the energy consumption of High Performance Computing (HPC) applications, and its characteristics i.e., power, area, latency, etc., is carried out. Until now an ideal platform for system-level evaluation is the gem5 simulator, as it provides methods for generating system environments with easily exchangeable separated components such as the memory controller, DRAM interface, and NVM interface.

### C. NOVEL CONTRIBUTION

In order to meet the requirement of designing a holistic system exploration framework, this paper introduces an innovative memory interface utilizing STT-RAM, making it a notable contribution. The interface has been created and seamlessly integrated into the gem5 simulator, establishing a connection with the included memory controller.

*The novel contributions of this paper are:*

- We propose HOPE which is an STT-RAM modeling and simulation framework integrated into the full system simulator gem5.
- We leverage the recently implemented NVM interface in gem5 to integrate HOPE with existing gem5 memory interfaces. This is in contrast to prior approaches that rely on external patches (like NVMain), which become less maintainable over time, thus stymying further development. Our proposed framework introduces a third memory interface tailored specifically for STT-RAM. This extension offers highly detailed results comparable to the existing DRAM implementation within gem5. Fortunately, integrating our framework into gem5 requires only minimal changes to gem5 files, as all functionality is implemented in new files that can be added seamlessly. Our implementation can be used identically to the existing memory interfaces and can potentially be integrated into the official gem5 repository by its core maintainers. Such integration would be the ideal outcome for our work, ensuring ongoing compatibility with gem5.
- We also extend the power model in gem5 DRAMPower to support our proposed STT-RAM model.
- We evaluate HOPE using HPC applications from the SPEC CPU 2017 benchmark suite on our event-driven gem5 simulator and successfully extract evaluation metrics from both the application and circuit perspectives.
- We will also open-source our framework to enable and accelerate the development of future system architectures based on STT-RAM.

The rest of the paper is organized as follows: Section II presents the various state-of-the-art works w.r.t. different types of memory controllers implemented inside gem5 and their drawbacks. In section III, the STT-RAM integration with the memory controller inside gem5 is discussed in detail. Section IV provides the evaluation metrics, results, and comparison. Finally, we conclude this paper in section V.

## II. BACKGROUND AND RELATED WORKS

STT-RAM has been recently exploited as an alternative to conventional on-chip memories because of its low energy consumption, high-speed access rate, scalability, and boundless endurance. However, several fundamental barriers, i.e., reliability issues due to Read/Write failure, Process Variation (PV) effects leading to stochastic switching time, should be considered before its vast industrial adaptation.

### A. OPERATION PRINCIPLES AND STT-RAM STRUCTURE

The most common structure of the STT-RAM cell includes an MTJ cell for data storage in series with an access transistor (1T1MTJ). MTJ cells include an oxide barrier sandwiched between two ferromagnetic layers called Rotation Layer (RL) and Fixed Layer (FL). The magnetization orientation between these two layers will result in two different states (i.e., parallel(P) and anti-parallel(AP)). These states are interpreted as an indicator of logic one and logic zero (see Fig. 1). The concept of Read and Write operation in the cell is explained in detail as follows:

#### 1) Write Operation

In order to write the intended information into the MTJ cell, a write current should be applied through the memory cell. A successful writing operation demands a minimum barrier exploiting the energy of the MTJ cell ($E_b$). If the required energy is supplied, then the state of the memory could be changed based on the current direction through the memory cell (see Fig. 1). In a coherent STT-RAM model, the required current to fulfill the required minimum barrier exploiting energy could be expressed as:

$$I_c = I_{c0}\left(1 - \frac{1}{\Delta}\ln(f_0 t_p)\right) \quad (1)$$

$$I_{c0} = \frac{8\alpha e M_s t}{\eta h \pi d^2} H_k \quad (2)$$

where, $I_{c0}$ : critical write current of the STT-RAM model (at 273 $^\circ$ Kelvin). This parameter is related to the physical property of the MTJ cell, such as: $M_s$: Material saturation magnetization; $f_0$: Attempt frequency which is typically ∼1 ns, $H_k$: Effective magnetic anisotropy field, $\alpha$: Damping factor, $\eta$: Spin polarization, $t_p$: Operating pulse-width (inverse of frequency), which provides an access frequency of 1 GHz; $t$, $d$: physical dimensions of the MTJ cell; $\Delta$: Thermal stability factor, and it can be expressed as:

$$\Delta = \frac{(M_s H_k t d)}{2 k_B T} \quad (3)$$

where, $k_B T$ : Describes the ambient energy in the system due to random thermal fluctuations [33].

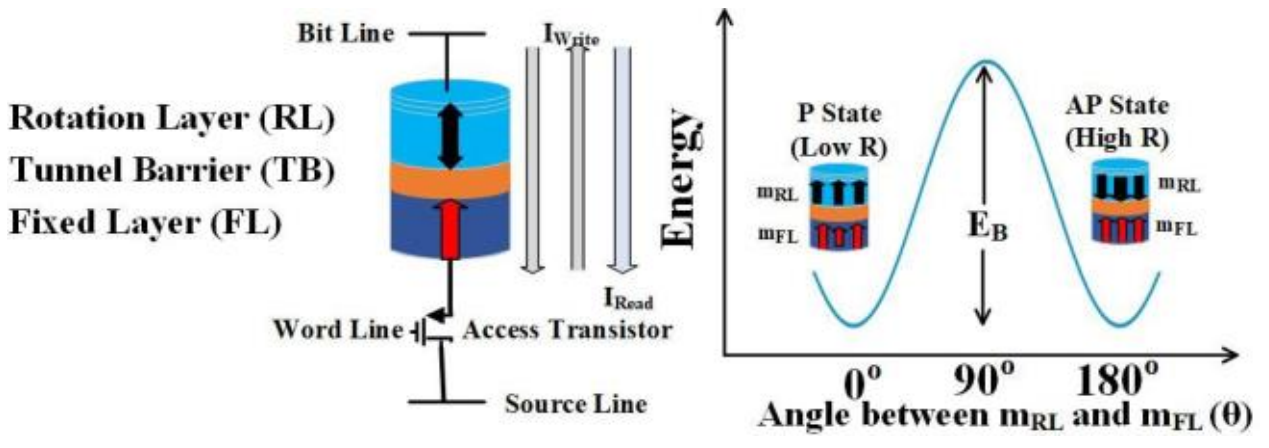

FIGURE 1: Schematic representation of magnetic orientation and energy barrier between two Magnetic Tunnel Junction (MTJ) states [33]

#### 2) Read Operation

Read operation in MTJ cell is due to the current $I_{read}$ passage through the cell. $I_{read}$ must be less than $I_{critical}$ and must not lead to change in the cell. The MTJ exhibits resistance value based on its two layers' magnetization (0 $^\circ$ or 180 $^\circ$ in Fig. 1). Therefore, by sensing, and passing the $I_{read}$ and measuring the resistance, we could identify the state of the cell. Moreover, the parameter Tunnel Magnetoresistance Ratio (TMR) is described as the difference between these two resistance states and can be expressed as $TMR = \frac{(R_{AP} - R_P)}{R_P}$, where $R_{AP}$ is the resistance in anti-parallel and $R_P$ is the resistance in parallel state. TMR is directly related to the cell's read latency, which means a higher TMR enables a faster and more precise read operation [33].

### B. RELATED WORK

Recent studies show NVM as the main memory element via a simulation in gem5, as gem5 was isolated from an NVM interface till October 2020. The lack of an NVM interface made it necessary for NVM research to use external tools such as NVSim [34], NVMain [35], and NVMain 2.0 [36]. To maintain compatibility, these NVM simulators must be developed simultaneously with the gem5 simulator. Especially NVMain offers a patch to be applied to the gem5 simulator, which connects the NVMain tool to the gem5 simulator. This patch directly modifies the gem5 source code and needs to be updated to match the latest gem5 releases. The result of the patch is a co-simulator where the gem5 simulator is in constant interaction with NVMain. Small changes in the gem5 simulator can directly require a change in the NVM simulation tools and the needed patches for integration into a co-simulator. The official patch for integrating NVMain into the gem5 was last updated in December 2016 and is incompatible with recent releases of the gem5. Furthermore, these state-of-the-art STT-RAM gem5 tool flow lacks available open-source models, which has hampered research and experimentation, impacting the adoption of STT-RAM in the current systems.

Therefore, in this manuscript, we propose a holistic gem5-based framework that will be open-sourced to fuel research and development in this area and further enhancement in the framework.

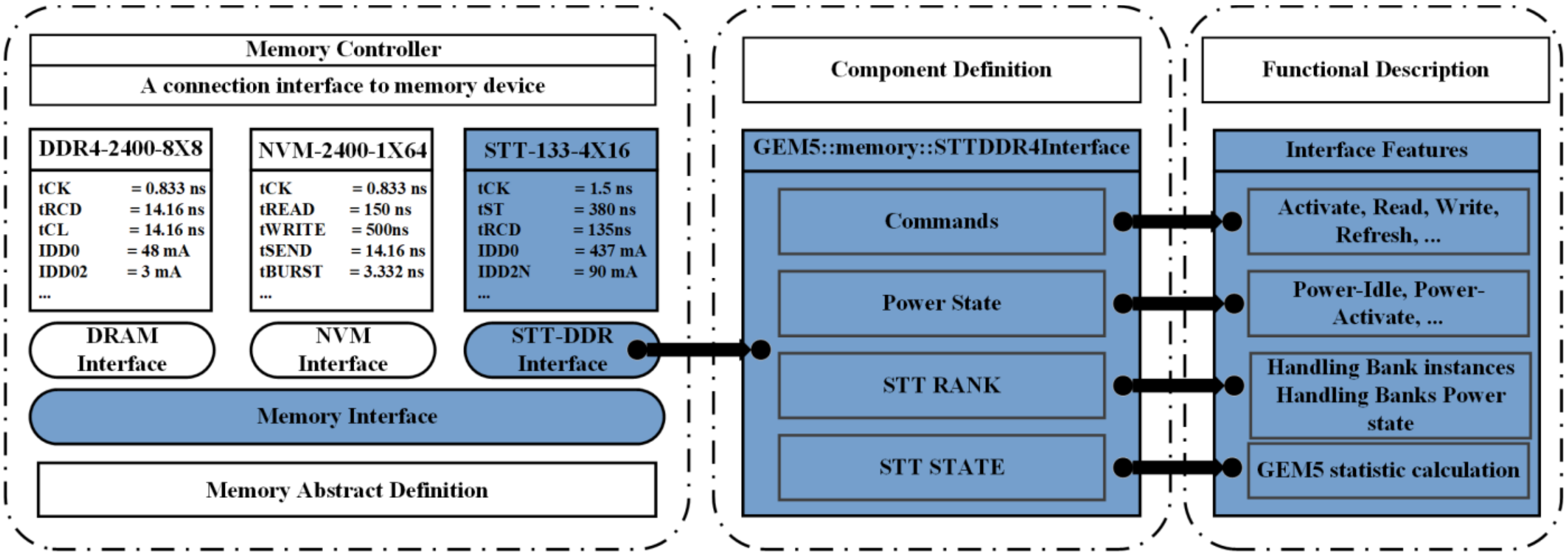

FIGURE 2: The STTDDR4 Interface integration into the gem5 standard components library. Components in blue are modified or new to gem5, and components in white are unmodified gem5 components.

The work presented in [37] explores some architecture-level modifications of STT-RAM structure aiming to provide NVM-based row buffers and reports a 67% energy improvement exploiting their approaches compared with state-of-the-art techniques. Moreover, authors in [29] explore the possibility of using STT-RAM to substitute the DRAM in main memory and evaluate their approach based on the SPEC CPU2006 dataset to be compared with DRAM-based memories. This study has been carried out on a trace-based cycle-accurate simulator. In [38], an STT-RAM-based memory has been proposed based on a 9F2-cell on the circuit level. The exploited MTJ model in this work requires a low switching current to change the state from logic-one to logic-zero. The application-level analysis has been estimated based on an HPC SPEC CPU 2017 benchmark for latency improvement. The aforementioned techniques either perform only circuit-level simulations (NVSim) which is typically time-consuming due to detailed hardware simulations, while other approaches that evaluate applications at the system level are not available open-source (to the best of our knowledge), thus obstructing the adoption of the STT-RAM model in the systems. *Our novel proposed HOPE framework has a fully integrated architectural model of STT-RAM in the gem5 simulator using a memory controller for exploiting system-level characteristics. This is in contrast to prior approaches that rely on external patches (like NVMain). HOPE is an event-driven gem5 simulator that facilitates system-level evaluation and enables the extraction of comparative metrics across all layers of the system hierarchy.*

## III. THE HOPE FRAMEWORK

### A. HOPE FRAMEWORK OVERVIEW

The novel proposed STT-RAM interface in this manuscript is an additional memory interface to the gem5. Therefore, in order to satisfy the compatibility of the STT-RAM interface with the gem5's MemCtrl component, there is a need for some modifications to the MemCtrl component. This tailoring has no effect on the existing functionality of connecting DRAM or NVM memories.

In the system configuration of gem5, the MemCtrl component offers a single port to connect a main memory instance to the MemCtrl, historically called *DRAM*, used for all types of memories. We introduced the STT-RAM interface to the gem5 as an alternative choice to the DRAMInterface and NVMInterface components as shown in Fig. 2. The interface is implemented in C++ (Component functional description), wrapped by a Python parameter configuration (part of Component definitions). The Python wrapper defines and inherits parameters that are mandatory for the component to work. The functionality of the STTDDR4Interface is placed within its C++ class. The component STT_1333_4x16 is the test device, populated with parameters from the EMD4E001G16G2 datasheet [39]. Our proposed framework enables access to modify the memory via its integrated interface and fetch the output data in our component definition. Thus, we integrated all the functionality into the STT-RAM memory controller through the provided interface to configure and evaluate the system analysis. This provides an edge over the state-of-the-art proposed methods that are based on the co-simulation of gem5 and other simulation tools (e.g., NVSIM, NVMAIN, etc.).

### B. HOPE STT-RAM POWER MODEL

The implemented STTDDR4 memory interface implementation represents a state machine consisting of states for idle, storing, activating, power up, and power down stages, as shown in Fig. 3. The MemCtrl instance of the simulated system and the memory logic help in the transition of the state in the state machine. As soon as the system is started, the very first state of the system is the PWR_IDLE state. This is the state from where the transition to activate state PWR_ACT, or activate with store state PWR_ACT_ST can be achieved using an ACT or ACT_ST command. The ACT_ST command is introduced to gem5 which is responsible for the STT-RAM-specific handling of data in the volatile page buffer.

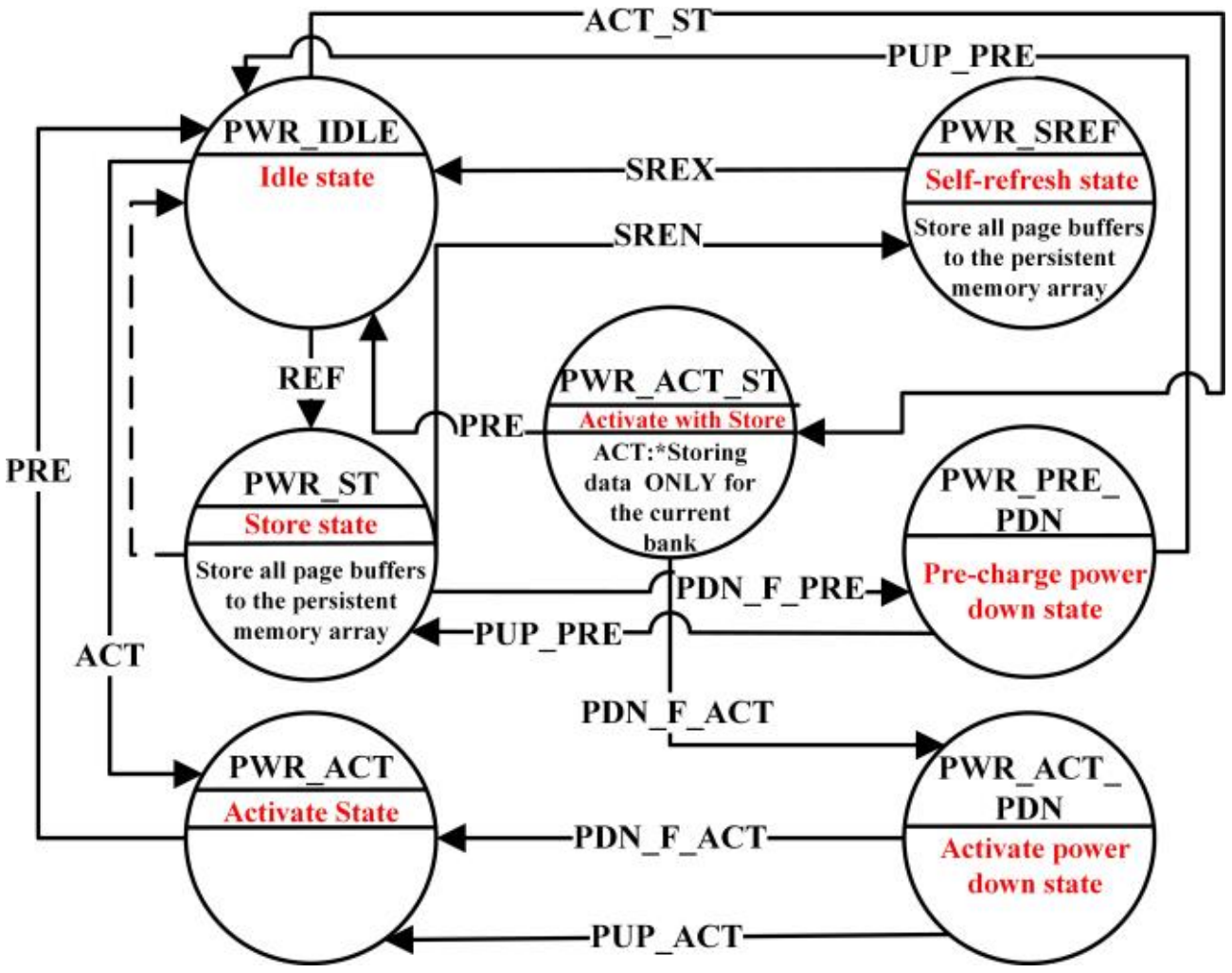

FIGURE 3: Power state machine of STTDDR4 integration to gem5

The main memory can be exploited using gem5 standard commands i.e., for a bank activation ACT (activate), RD (read), WR (write), REF (refresh all banks), SREF (Self-Refresh), PRE (explicit pre-charge of a single bank), etc. In this manuscript, an additional command is been introduced i.e., ACT_ST (activate with the store). In the EMD4E001G16G2 datasheet, [39], the introduced ACT_ST command is known as ACT*. The ACT* command includes a store procedure for the specific bank accessed. The implementation of the ACT_ST command to gem5 also includes a new event, the *actStoreEvent*, to enable the transition to the new power state PWR_ACT_ST. Moreover, the automatic execution of REF commands needed for data persistence in DRAM is not included in STTDDR4, as refreshes are not mandatory in STT-RAM devices as per the EMD4E001G16G2 devices.

As we know that method calls are responsible for MemCtrl interaction with the memory interface during simulation

**Algorithm 1** Select between ACT and ACT_ST command on rank:bank:row

bank.storingState ← PERSISTENT
bank.lastRow ← 0

**procedure** ActivateBank($rank, bank, row$)
  $cmd \leftarrow ACT$
  **if** $bank.lastRow \neq row$ **then**
    **if** $bank.storingState = BUFFER$ **then**
      $cmd \leftarrow ACT_ST$
      $cmdDelay \leftarrow cmdDelay + tST$
    **end if**
  **end if**
  $cmdList.push_back(cmd, bank, delay)$
  $bank.lastRow \leftarrow row$
  $bank.storingState \leftarrow BUFFER$
  **process** $cmd$ in drampower

**Algorithm 2** DRAM Power Extension

nOfActsBanks[banks] ← zeros(banks)
nOfStoresBanks[banks] ← zeros(banks)

**procedure** EvaluateCommands(*cmdList*)
  **for all** $cmd \in cmdList$ **do**
    **if** $cmd.type = ACT$ **then**
      $handleAct(cmd.bank)$
    **else if** $cmd.type = ACT_ST$ **then**
      $handleActSt(cmd.bank)$
    **else if** <other command types> **then**
      <handle commands>
    **end if**
  **end for**
**end procedure**

**procedure** HandleActSt(*bank*)
  **if** isPrecharged(bank) **then**
    $nOfActsBanks[bank] ++;$
    $nOfStoresBanks[bank] ++;$
  **end if**
**end procedure**

**procedure** power_calc()
  $calc(sum(nOfStoresBanks) * tST, idd0 - idd3n)$
  **for all** $bank \in banks$ **do**
    $calc(nOfStoresBanks[bank] * tST, idd0 - ione)$
  **end for**
**end procedure**

**procedure** calc(*cycles*, *current*)
  **return** (cycles*clkPeriod*current*voltage)
**end procedure**

e.g., while reading or writing data, the MemCtrl initiates a burst access to the memory device where MemCtrl provides data for which rank, bank and row the burst access needs to be executed. This rank, bank, and row information are handed over to the bank activation method, as shown in algorithm 1. The EMD4E001G16G2 device includes functionality for automatic storing of page memory data to the persistent memory array when the page memory data would become lost otherwise. The MemCtrl does not offer functionality for differentiation of the storing states in STT-RAM.

Therefore, the STTDDR4Interface got extended with functionalities to track the storing state of the data in the page buffer of each bank. The storing state of each bank supports the states BUFFER and PERSISTENT. All the banks during startup are initialized with PERSISTENT, which indicates the page buffer data to be saved to the persistent memory array. Moreover, the state BUFFER proves the bank to be not saved to the bank's persistent memory array. Also, the last row accessed will be saved in each bank and updated during each bank activation. In order to change the storing state of a bank, or all banks, to PERSISTENT, a store operation needs to be performed. This can be triggered by an ACT_ST command, a REF command, or an SREF command. Within a REF or SREF execution, store operations on all banks in storing state BUFFER will be performed. When there are no banks in storing state BUFFER, the commands REF and SREF are not effective.

The selection between the ACT or ACT_ST command is done in the activated bank method with a sequence of procedure calls as Algorithm 1. The requested row is compared to the last accessed row of the particular bank. When the last accessed data is still the working data, which means the last access row and requested row are identical, there is no need for a store operation. In this case, a normal ACT command will be simulated. If the requested row differs from the last accessed row, and the bank is in the storing state ”BUFFER“, an ACT_ST command will be pushed to simulate. The difference in simulating ACT or ACT_ST is implemented in a higher delay for the additional store operation, which is known as the store time (tST). This store procedure call impacts the energy consumption that can be calculated in the power library. The bank's storing state can be changed to ”PERSISTENT“ by performing a store operation on the particular bank, or on all banks. The ACT command is simulated as with DRAM memory in gem5: The ACT command is saved in a list of pending commands which is handed over to the modified DRAMPower tool [40], which is part of gem5 and performs energy calculations from the gem5 inputs.

**Extensions to DRAMPower Model:** To be able to include the ACT_ST command, DRAMPower got extended by the command and energy calculation. The extensions to DRAMPower are presented in Algorithm 2. These extensions include functionality for counting the number of store procedures during runtime and calculating the resulting store energy and power.

Moreover, the energy calculation in gem5 is not updated on every new command execution, but on specific simulation phases: suspension, end of a refresh (REF) command, triggered from the gem5 system configuration script or by the internal command line of the full system simulation using the m5 utility (the gem5 utility used in full system disk images).

Furthermore, the gem5 statistic output has been modified to include the store energy per rank and power state time per rank in the simulation results. In this section, we present our HOPE framework for an all-integrated STT-RAM with a gem5 simulator using the memory controller. This helps in exploiting system-level meta-heuristics that include power consumption, memory utilization, heat patterns, etc.

### C. HOPE CONFIGURATION

The gem5 being an instruction-level simulator enables the integration of different types of memories with a memory controller. The memory controller is a component that enables an option of choosing memory for system-level analysis. The memory controller has evolved a lot in the past few years. Recently, in May 2020 the gem5 introduced a new memory controller (MemCtrl) component revision and introduced an NVM interface (NVMInterface) class to the gem5, officially integrated into version 20.1 of the gem5 simulator. This NVM interface is designed very generic in terms of its functionality and parameters to be taken into consideration. The NVMInterface class offers three timing parameters: tREAD, tWRITE, and tSEND.

TABLE 2: The Configuration of the Memory cell

| Paramters | Configuration |
|---|---|
| Memory | 1Gbit x16 |
| Organization | 8-banks (2-banks per Bank Group) |
| Latency | 667MHz (1333MT/s) |
| Access Time | 225ps |
| Supply Voltage - Min: | 1.14 V |
| Supply Voltage - Max: | 1.26 V |

There is also an already existing DRAM interface (DRAMInterface) class. This class of the gem5 contains detailed logic on DRAM timing and power state simulation and offers various timing, e.g. tRCD, tCL, tRAS, and tREFI, and energy parameters, e.g. IDD0, IDD4R, and VDD. But there is no such logic for calculating NVM energy and power consumption. Also, there are no energy parameters available for NVMInterface.

Thus, to overcome such shortcomings, HOPE provides another detailed memory interface targeting STT-RAM. This memory targets real-world STT-RAM devices which are designed as STT on top of DDR4. Therefore, this interface is named as STTDDR4Interface. This offers a high level of detail timing and energy parameters, combined with a power state and energy calculation logic. Fig. 2 depicts the detailed flow of our HOPE framework within the extended and modified gem5 simulator.

The simulated system is configured using the fs.py system configuration script. Using this script, a System instance is set up according to the input values of the HOPE framework. We use an X86 architecture-based system. The CPU we defined is the TimingCPU, which offers detailed memory timings during simulation. The CPU is also equipped with L1 and L2 caches. The system uses a Linux kernel and a disk image with an Ubuntu operating system and workloads installed. Within gem5, communication between system blocks is done via ports, also as real systems do. The system block is connected to the Membus block. The memory bus selected is by default, the SystemXBar. All CPU interactions to the main memory are forwarded by the Membus to the memory controller. MemCtrl got modified to support STT-RAM connected through the memory port.

Fig. 2, shows the proposed architecture with the STT_1333_4x16 which is a class created for interfacing STT-RAM with the memory controller. It has multiple parameters e.g., tCK, tST, tRCD, IDD0, IDD2N, etc that has been extracted from the datasheet of the aforementioned device.

The tCK is the clock period, depending on the device operating clock frequency (fCK) (e.g., fCK = 667MHz results in 1.5ns tCK (=1/fCK)), tST is a special timing parameter for STT-RAM and refers to the storing time of the memory (indicates the time needed for storing the data from the row address buffer to the persistent STT memory). The tST is a newly introduced parameter to gem5 unique to STT, which was a missing timing parameter for gem5. The address buffer acts like a cache, and the data placed in the cache needs to be written to the main STT memory during the Store operation. Therefore, tST is the time needed to process data moving from the row address buffer to the STT persistent memory array. Researchers in the field could optimize different metrics to minimize this value and evaluate the performance of so-called in-memory processing approaches. The parameters such as tRCD, IDD0, IDD2N, etc., are standardized DDR4 parameters.

TABLE 3: Experimental systems configuration for STT-RAM and DRAM

| System elements | Processor | L1 Instr. cache | L1 Data cache | L2 cache | Main Memory | Clock Speed | Row Buffer Size | Device Size | Channel Cappacity | tRCDmin | tRCmin | tRASmin | tFAWmin | tRPmin |
|---|---|---|---|---|---|---|---|---|---|---|---|---|---|---|
| **STT-RAM** | 64-bit x86 single core, timing 3GHz | Private, 32kB | Private, 32kB | Shared, 256kB | 1 channel, 2 rank per channel, 4 chips per rank, EMD4E001G16G2, 1Gbit x16, 1333MHz | 667MHz | 256B | 128MiB | 1GiB | 135ns | 44.5ns | 32ns | 15ns | 12.5ns |
| **DRAM** | " | " | " | " | 1 channel, 2 rank per channel, 8 chips per rank, MT40A1G8SA, 1Gbit x8, 2400MHz | 1200MHz | 1kB | 1GiB | 16GiB | 12.5ns | 190ns | 143ns | 240ns | 7.5ns |

Moreover, this interface makes it possible to simulate systems using the latest STT-RAM devices including power states and energy consumption as it was never possible before in a stand-alone gem5 environment. The integration carried out on the interface is based on the parameters offered by the STT-RAM EMD4E001G16G2 from Everspin Technologies [39]. These device parameters are shown in Table 2.

As per the physical characteristics of STT-RAM, there are deviations to the DDR4 specification for DRAM especially the *Refresh* command, which is mandatory to be issued in a time interval tREFI on DRAM, is no longer used in STT-RAM. Therefore, tREFI got removed for STT memory. Moreover, the STT-RAM also has a store time parameter tST. The store operation of delay tST, is used to move recently written data from the page buffer to the persistent memory array.

Some other deviations specific to the test devices (simulating the EMD4E001G16G2 device) include the memory size that in the case of STT-RAM is a 1 Gbit device, whereas the DDR4 specification for DRAM only allows devices of 2, 4, 8 and 16 GBit. Furthermore, there is also a limit of 667 Mhz for the clock frequency, while the DDR4 Specification for DRAM allows 800, 933, 1067, and 1200MHz.

### D. EVALUATION SETUP CONFIGURATION

Fig. 4 presents a comprehensive overview of the HOPE framework setup configuration steps, highlighting its key contributions depicted in blue. gem5 full system simulations require a disk image prepared with an operating system and a kernel compatible with the chosen operating system. The HOPE framework uses a 1 modified Packer SPEC CPU 2017 setup script from the gem5 resources repository for generating a disk image containing the Ubuntu operating system and the SPEC CPU 2017 benchmark suite for X86 architecture. The 2 benchmark installation is then followed finalized by mounting the disk image on the host system. Each benchmark from the disk image has been run once for finalizing the benchmark installation, this includes compiling, training, and running the benchmark. In 3 the gem5 full system simulation including the HOPE extensions and modifications is run. Therefore the created disk image is used. Each simulation runs a selected workload from the prepared disk image, includes the creation of checkpoints after the operating system boot, and the output of the 4 detailed statistics after the gem5 simulation is completed. The gem5 full system simulation includes the introduced STT-RAM extension and modifications to DRAMPower to allow detailed energy calculation for our STT-RAM device. 5 shows the provided McPAT template file, modified to support the extended outputs of gem5. Using the system configuration, simulation statistics, and McPAT template, the 6 the GEM5ToMcPAT [41] tool is used to generate an input file for later use with McPAT. HOPE includes the enhanced 7 cMcPAT power, area, and timing modeling framework. cMcPAT [41] is capable of calculating the power parameters of 9. Using the statistics output of gem5, and the power model of cMcPAT, the script ”print_energy“ [41] is calculating the total energy consumption of the simulated environment. The results also combine the detailed output of gem5, especially, the instructions count.

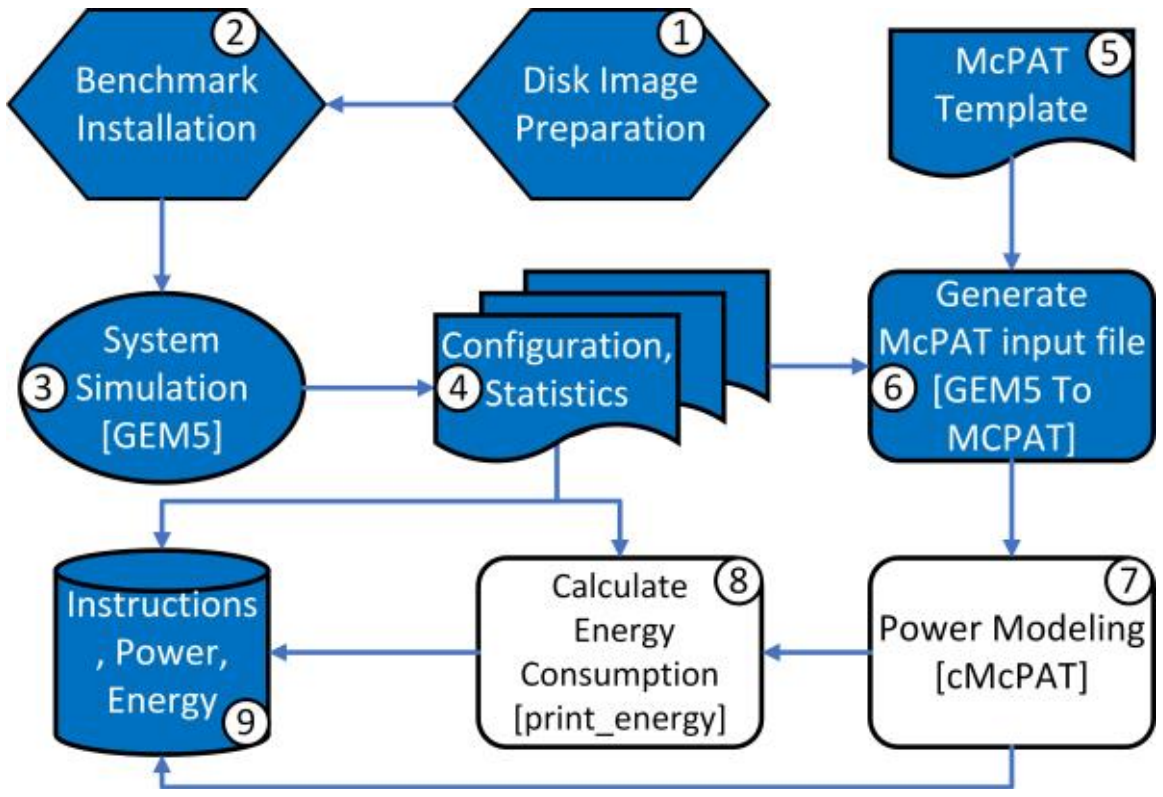

FIGURE 4: Overview of HOPE framework

The simulated system is configured using the gem5 stdlib (gem5 standard library) based on an X86 configuration. Table 3 lists the detailed configuration of the processor, cache, and memory for both experimental systems using STT-RAM and DRAM. We selected an existing STT-RAM device for simulation and paired it with a widely used DRAM device to facilitate a functional comparison. It's important to note that our chosen STT-RAM device does not align with JEDEC's JESD79-4A DDR4 standard, which limits our ability to select a DDR4-based DRAM device with nearly identical parameters.

Moreover, Fig. 5 shows the system configuration used for benchmarks using STT-RAM. gem5 offers a general full system default configuration script (fs.py), which we used in this work to reflect the system architecture of our simulated system. We used the gem5 full system emulation mode to reflect real-world systems in the best way provided.

The operating system selected is an Ubuntu 18.04 configured for gem5 and set up with a SPEC CPU 2017 benchmark suite [42] instance. The kernel used is the linux kernel version 4.19.83. The gem5 configuration script handles the creation of memory controllers and memory devices. The count of memory ranks and banks is set in the memory device configuration. The parameters used for the STT-RAM device configuration in gem5 are sourced from the EMD4E001G16G2 datasheet [39]. We performed SPEC CPU 2017 benchmarks on our simulated systems using 2 checkpoints per benchmark to be able to perform detailed simulations using the TimingCPU from gem5.

The first checkpoint has been saved after the OS boot is finished. The second checkpoint has been saved after the first 4.5 billion instructions of benchmark application to ensure the initialization phase has been finished and the checkpoint defines a direct jump into the main benchmark algorithm. Both checkpoints were performed in a fast-forward method using gem5's AtomicSimpleCPU. The main simulation run was done from the second checkpoint for a total of 2 Billion instructions. This procedure has been performed for all of the benchmark applications included in SPEC CPU 2017.

## IV. RESULTS

In our research endeavour, our primary focus revolved around conducting simulations utilizing the parameters of real-world devices. Specifically, we honed in on a selected STT-RAM device, which holds the distinction of being DDR4-compatible, albeit with certain deviations from the official standard. As we delved into the simulations, we observed that these deviations had a tangible impact on our results, which we duly documented.

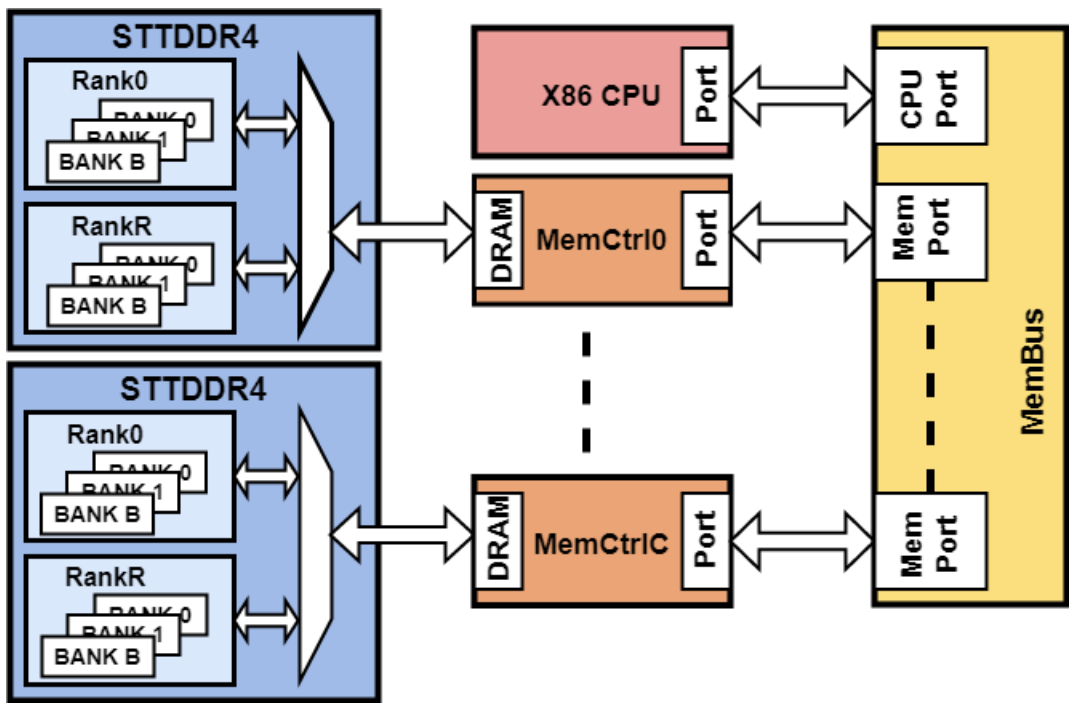

FIGURE 5: Architecture of the STT-RAM simulated system.

Our findings shed light on the applicability of HOPE as a potent tool for evaluating the feasibility of incorporating STT-RAM main memories into practical systems. By harnessing the capabilities of HOPE, we were able to gain insightful glimpses into the performance and energy efficiency of the SPEC CPU 2017 benchmarks when paired with STT-RAM. This analytical approach not only offers a valuable lens to understand the potential of STT-RAM but also opens up new vistas of exploration and optimization in memory technology research.

The gem5 simulator provides users with multiple avenues for creating system configurations through its gem5 standard library of components and functions, including CPUs, memory, boards, cache hierarchies, and more. This comprehensive collection of building blocks is commonly referred to as the gem5 standard library. For instance, if you wish to modify the CPU architecture, you can simply select the corresponding board (e.g., X86Board or ArmBoard) and fine-tune the memory configurations accordingly. This flexible approach can be applied to alter any available system component.

Alternatively, there is an option to configure a system within gem5 without utilizing a predefined board. Instead, you can manually establish connections between a CPU and a selected memory device using a memory bus. Furthermore, you have the flexibility to augment the CPU with either a straightforward single-level cache or a more intricate cache hierarchy to suit your needs.

Our proposed framework takes a distinct approach by leveraging an existing system configuration known as fs.py. This configuration can be effortlessly modified via command-line inputs, enabling rapid adjustments to the system configuration with a single bash script edit. Different fs.py configurations are available for various system architectures, such as *configs/example/riscv/fs_linux.py* for RISC-V or *configs/example/arm/fs_bigLITTLE.py* for ARM.

In our research endeavors, we conducted extensive simulations employing both fs.py and custom gem5 system configuration scripts. This comprehensive approach allowed us to thoroughly assess and analyze our simulations.

### A. STT-RAM STATE TIME DISTRIBUTION

As depicted in Fig. 8(b), the power state times for the IDLE state exhibit variations in the case of STT-RAM, contingent upon the specific workload. Notably, STT-RAM distinguishes itself by eschewing the need for periodic refreshes to maintain data states, leading to a complete absence of time spent in the REF power state. Conversely, the bank activation (ACT) time, also illustrated in Fig. 8(b), demonstrates only a minor variation within the STT-RAM-based system. This effect can be attributed to the relatively prolonged delays observed in the bank activation process, especially concerning store state (ACT_ST) for STT-RAM, as visually demonstrated in Fig. 8(b). Such insights garnered from our analyses provide valuable perspectives on the dynamic behaviour and efficiency of STT-RAM, imparting crucial knowledge for potential real-world implementations and optimizations in-memory technologies.

### B. STT-RAM PERFORMANCE EVALUATION

Fig. 6(a) show the total amount of read and write requests which are generally higher in the STT-RAM devices. High read and write requests are explained by Fig. 8(a) which shows the hit rate for read operations. The hit rates for STT-RAM depend on the application complexity level. Fig. 8(c) shows the average read and write bandwidth with a simulated STT-RAM. Also, the average latency of STT-RAM, shown in Fig. 6(b), for each application highly depends on the hit ratio. As described in our STTDDR4 power state machine description, a high alteration of accessed rows negatively affects the energy and latency of our simulated STT-RAM device. A high alteration of accessed rows further has an impact on store operations.

### C. STT-RAM POWER AND ENERGY BREAKDOWN

In Fig. 6 (c and d), we present a comprehensive view of the power and energy breakdown for our simulated systems, offering valuable insights into their performance characteristics. The shown parameters in d represent the accumulated energy for different commands issued to the memory device. As shown in Fig. 3, the memory devices move through different states during runtime. The parameter ”Activation energy“ therefore shows the total energy consumption for all ACT commands initiated. ”Store energy“ accumulates the energy of all store operations during the evaluation. In the case of an ACT_ST operation, the energy accumulated separately for the activation and the store energy results. The parameter ”Refresh energy“ is associated to the REF command, while the ”RD/WR energy“ is the accumulated energy during the processing of the read and write burst accesses.

Particularly noteworthy is the substantial count of store operations, which, in conjunction with the notably high IDD0 current of 437mA, prominently influences the calculated store energy. As evident from the results, STT-RAM stands out for its lack of refresh energy requirements. However, it should be noted that the stored energy demands contribute to an overall increase in the total energy breakdown. These findings shed light on the contrasting energy consumption patterns of STT-RAM compared to conventional memory technologies, signifying the potential for more energy-efficient computing paradigms. The comprehensive understanding gained from these power and energy analyses is essential for devising strategies to optimize memory architectures, thus fostering advancements in the realm of energy-efficient computing systems. The presented energy parameters in d are not a full view of all calculated energy parameters within gem5, but an excerpt of significant values. The full list of energy parameters is extended by parameters for interface energy, standby energy, self-refresh energy, power-down and power-up energy.

### D. DRAM METRICS

To maintain the extendibility and versatility of our framework, we have thoughtfully retained the interface to the DRAM. This strategic decision allows our framework to adapt effortlessly to various memory technologies, rendering it highly versatile for a wide array of computing scenarios. In this section, we present an in-depth analysis of the extracted data concerning the state time distribution, power consumption, energy usage, and latency breakdown for the same applications from SPEC 2017, with the DRAM serving as the primary memory. This comprehensive investigation is instrumental in understanding the behavior and performance characteristics of our framework when interfacing with DRAM.

Fig. 9(b) illustrates the state time distribution of these applications when utilizing DRAM. Additionally, Fig. 7(a) showcases the framework's memory requests. Analyzing memory requests offers deeper insights into the applications´ memory access patterns, shedding light on potential areas for improvement in terms of data locality and cache utilization. Fig. 9(c) provides insight into bandwidth usage with DRAM as the memory. Bandwidth utilization is a critical metric for assessing memory system efficiency and identifying potential bottlenecks that may impact application performance. Furthermore, Fig. 9(a) reveals DRAM row hits, and Fig. 7(b) presents latency per application. Finally, Fig. 7(c and d) exhibits DRAM average power and energy usage while running the SPEC 2017 applications. These detailed analyses offer valuable insights into our framework's performance and its potential for adaptation to future memory technologies and diverse computing environments.

### E. COMPARING STT-RAM AND DRAM RESULTS

Based on our evaluation of SPEC 2017 benchmarks, it becomes evident that STT-RAM is not yet ready to replace DRAM-based main memories for many applications due to its higher store latency and energy consumption. STT-RAM needs storing from the page buffer to the persistent memory array whereas DRAM does not need this. Due to the overhead for storing data in the STT persistent memory array within the ACT-ST state, the delays are significantly higher than in DRAM. Each store takes 380ns extra, in all cases of ACT-ST state. Furthermore, STT-RAM is running at 1333 MHz, whereas DRAM is running at 2400 MHz. The impact is especially pronounced in applications with high write-to-read ratios like Ibm_s (see Fig. 6a). Before STT-RAM can be a feasible alternative to DRAM's main memories, further technology, and architectural optimizations are necessary to reduce the store latency and energy requirements. Fortunately, with the availability of HOPE, we now have a systematic means to evaluate and optimize STT-RAM at the system level. HOPE presents an invaluable opportunity to drive STT-RAM's progress by allowing us to

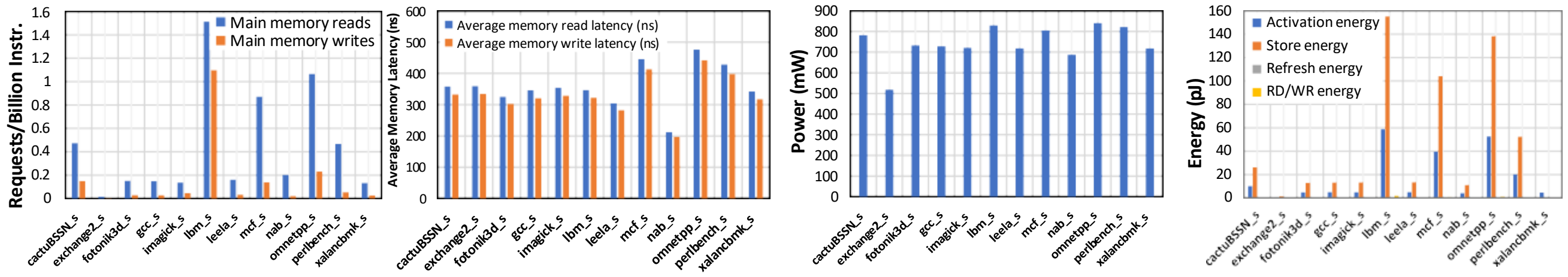

(a) R/W requests/billio. instr. (b) Average memory latency(ns) (c) Power consumption(mW) (d) Energy usage (pJ)

FIGURE 6: STT-RAM evaluation metrics for different application

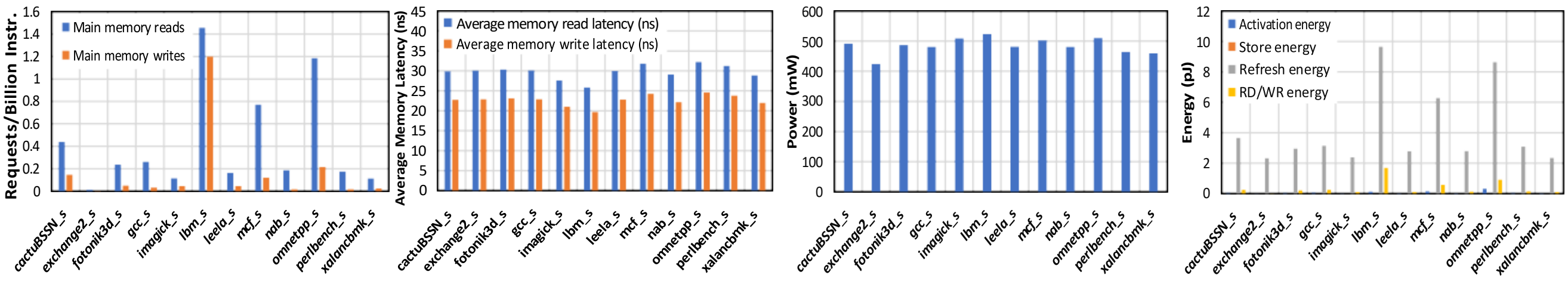

(a) R/W requests/billio. instr. (b) Average memory latency(ns) (c) Power consumption(mW) (d) Energy usage (pJ)

FIGURE 7: DRAM evaluation metrics for different application

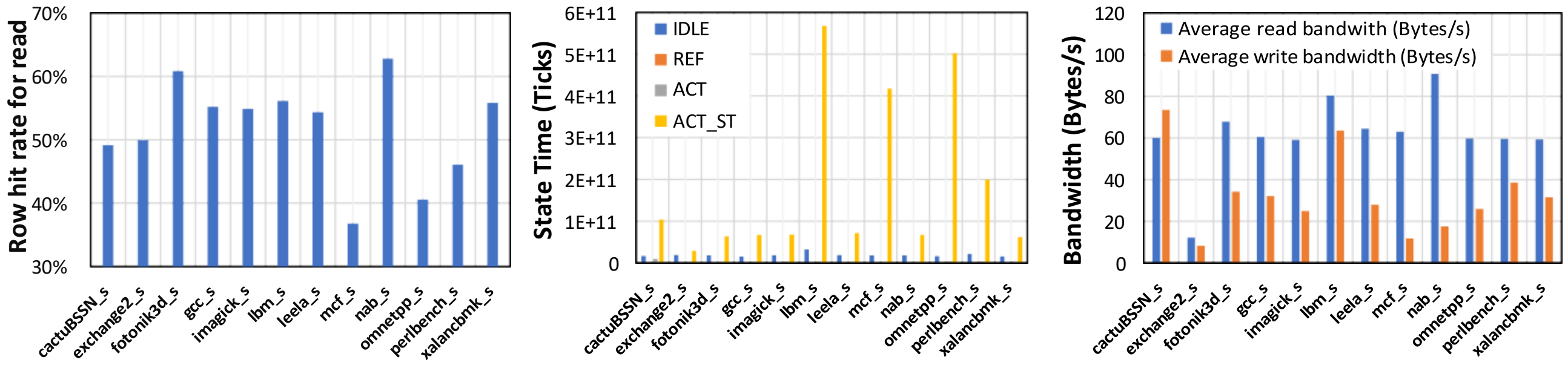

(a) Read operation row hits(%) (b) State delays (# ticks) (c) Average R/W bandwidth (Bytes/sec)

FIGURE 8: STT-RAM row hits, state time, and bandwidth usage

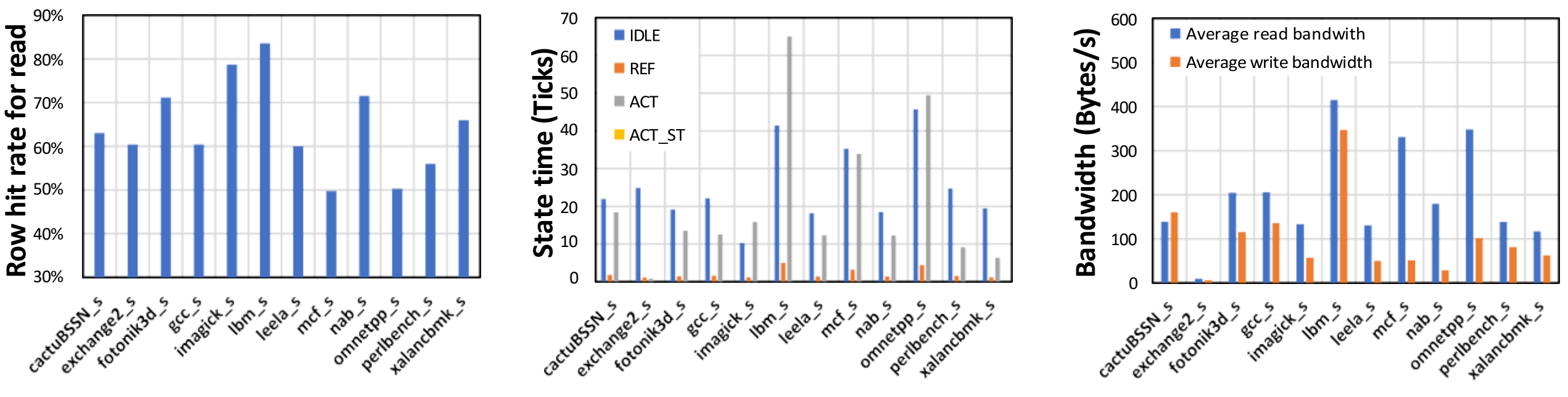

(a) Read operation row hits(%) (b) State delays (# ticks) (c) Average R/W bandwidth (Bytes/sec)

FIGURE 9: DRAM row hits, state time, and bandwidth usage

explore and analyze various architectural possibilities with greater precision.

## V. CONCLUSION

We presented an extension to the open-source full-system simulator gem5 for enabling detailed evaluation of STT-RAM devices in an accurate manner. We have shown our implemented power state machine, memory commands, power calculation, and statistics output. We have also shown the results of an STT-RAM-based system configured using real-world device parameters and compared the resulting metrics to a DRAM-based system. The STT-RAM-specific characteristics of required store operations and deviations to the DDR4 standard for DRAM have been discussed based on the comparison of the simulation results. Our HOPE implementation is easily configurable for other STT-RAM devices, by adding timing values, currents, and voltages. We will open-source our HOPE framework to fuel research and accelerate the development of future system architectures based on STT-RAM.

## VI. ACKNOWLEDGMENT

The authors acknowledge TU Wien Bibliothek for financial support through its Open Access Funding Programme.

## REFERENCES

[1] Y. Wang, C. Zhang, H. Yu, and W. Zhang, “Design of low power 3d hybrid memory by non-volatile cbram-crossbar with block-level data-retention,” in Proceedings of the 2012 ACM/IEEE international symposium on Low power electronics and design, 2012, pp. 197–202.

[2] Y. Shin, “Non-volatile memory technologies for beyond 2010,” in Digest of Technical Papers. 2005 Symposium on VLSI Circuits, 2005. IEEE, 2005, pp. 156–159.

[3] B. Gervasi, “Will carbon nanotube memory replace dram?” IEEE Micro, vol. 39, no. 2, pp. 45–51, 2019.

[4] J. Lamb, S. Gibbons, R. Trichur, Y. Jiang, K. Mangelson, K. Kremer, and D. Janzen, “Advancements in microelectronics-grade carbon nanotube materials for nram® device manufacture and analysis of carbon nanotube mass in end user devices.”

[5] “Irds 2022 beyond cmos and emerging materials integration,” Online, 2022, accessed on: November 7, 2023. [Online]. Available: https://irds.ieee.org/editions/2022

[6] D. Jana, S. Roy, R. Panja, M. Dutta, S. Z. Rahaman, R. Mahapatra, and S. Maikap, “Conductive-bridging random access memory: challenges and opportunity for 3d architecture,” Nanoscale research letters, vol. 10, pp. 1–23, 2015.

[7] S. Mittal and J. S. Vetter, “A survey of software techniques for using non-volatile memories for storage and main memory systems,” IEEE Transactions on Parallel and Distributed Systems, vol. 27, no. 5, pp. 1537–1550, 2015.

[8] J. S. Meena, S. M. Sze, U. Chand, and T.-Y. Tseng, “Overview of emerging nonvolatile memory technologies,” Nanoscale research letters, vol. 9, pp. 1–33, 2014.

[9] T. Mikolajick, C. Dehm, W. Hartner, I. Kasko, M. Kastner, N. Nagel, M. Moert, and C. Mazure, “Feram technology for high density applications,” Microelectronics Reliability, vol. 41, no. 7, pp. 947–950, 2001.

[10] M. Imani, S. Patil, and T. Rosing, “Low power data-aware stt-ram based hybrid cache architecture,” in 2016 17th international symposium on quality electronic design (isqed). IEEE, 2016, pp. 88–94.

[11] S. Jeloka, Z. Wang, R. Xie, S. Khanna, S. Bartling, D. Sylvester, and D. Blaauw, “Energy efficient adiabatic fram with 0.99 pj/bit write for iot applications,” in 2018 IEEE symposium on VLSI circuits. IEEE, 2018, pp. 85–86.

[12] M. Moore et al., “International roadmap for devices and systems,” Accessed: Jan, 2020.

[13] I. Yoon, A. Anwar, T. Rakshit, and A. Raychowdhury, “Transfer and online reinforcement learning in stt-mram based embedded systems for autonomous drones,” in 2019 Design, Automation & Test in Europe Conference & Exhibition (DATE). IEEE, 2019, pp. 1489–1494.

[14] B. Narasimham, V. Chaudhary, M. Smith, L. Tsau, D. Ball, and B. Bhuva, “Scaling trends in the soft error rate of srams from planar to 5-nm finfet,” in 2021 IEEE International Reliability Physics Symposium (IRPS). IEEE, 2021, pp. 1–5.

[15] J. Wang, N. Xiu, J. Wu, Y. Chen, Y. Sun, H. Yang, V. Narayanan, S. George, and X. Li, “An 8t/cell fefet-based nonvolatile sram with improved density and sub-fj backup and restore energy,” in 2022 IEEE International Symposium on Circuits and Systems (ISCAS), 2022, pp. 3408–3412.

[16] J. Y. Kim, M.-J. Choi, and H. W. Jang, “Ferroelectric field effect transistors: Progress and perspective,” APL Materials, vol. 9, no. 2, p. 021102, 02 2021.

[17] S. Yu, Q. Wang, Y. Zhang, P. Yang, X. Luo, H. Liu, C. Chen, Q. Li, and S. Liu, “Multistate capability improvement of beol compatible fefet by introducing an al2o3 interlayer,” IEEE Transactions on Electron Devices, vol. 70, no. 11, pp. 5632–5637, 2023.

[18] J. Y. Park, D.-H. Choe, D. H. Lee, G. T. Yu, K. Yang, S. H. Kim, G. H. Park, S.-G. Nam, H. J. Lee, S. Jo, B. J. Kuh, D. Ha, Y. Kim, J. Heo, and M. H. Park, “Revival of ferroelectric memories based on emerging fluorite-structured ferroelectrics,” Advanced Materials, vol. 35, no. 43, p. 2204904, 2023.

[19] S. Seyedfaraji, J. T. Daryani, M. M. S. Aly, and S. Rehman, “Extent: Enabling approximation-oriented energy efficient stt-ram write circuit,” IEEE Access, vol. 10, pp. 82 144–82 155, 2022.

[20] S. M. Nair, R. Bishnoi, A. Vijayan, and M. B. Tahoori, “Dynamic faults based hardware trojan design in stt-mram,” in 2020 Design, Automation & Test in Europe Conference & Exhibition (DATE). IEEE, 2020, pp. 933–938.

[21] R. Bishnoi, M. Ebrahimi, F. Oboril, and M. B. Tahoori, “Improving write performance for stt-mram,” IEEE Transactions on Magnetics, vol. 52, no. 8, pp. 1–11, 2016.

[22] S. Swami and K. Mohanram, “Reliable nonvolatile memories: Techniques and measures,” IEEE Design & Test, vol. 34, no. 3, pp. 31–41, 2017.

[23] S. Seyedfaraji, A. M. Hajisadeghi, J. Talafy, and H. R. Zarandi, “Dysco: Dynamic stepper current injector to improve write performance in stt-ram memories,” Microprocessors and Microsystems, vol. 73, p. 102963, 2020.

[24] E. Garzon, R. De Rose, F. Crupi, L. Trojman, G. Finocchio, M. Carpentieri, and M. Lanuzza, “Assessment of stt-mrams based on double-barrier mtjs for cache applications by means of a device-to-system level simulation framework,” Integration, vol. 71, pp. 56–69, 2020.

[25] R. Saha, Y. P. Pundir, and P. K. Pal, “Design of an area and energy-efficient last-level cache memory using stt-mram,” Journal of Magnetism and Magnetic Materials, vol. 529, p. 167882, 2021.

[26] E. Cheshmikhani, H. Farbeh, and H. Asadi, “3rset: Read disturbance rate reduction in stt-mram caches by selective tag comparison,” IEEE Transactions on Computers, vol. 71, no. 6, pp. 1305–1319, 2021.

[27] ——, “Robin: Incremental oblique interleaved ecc for reliability improvement in stt-mram caches,” in Proceedings of the 24th Asia and South Pacific Design Automation Conference, 2019, pp. 173–178.

[28] N. Mahdavi, F. Razaghian, and H. Farbeh, “Data block manipulation for error rate reduction in stt-mram based main memory,” The Journal of Supercomputing, vol. 78, no. 11, pp. 13 342–13 372, 2022.

[29] E. Kültürsay, M. Kandemir, A. Sivasubramaniam, and O. Mutlu, “Evaluating stt-ram as an energy-efficient main memory alternative,” in 2013 IEEE International Symposium on Performance Analysis of Systems and Software (ISPASS), 2013, pp. 256–267.

[30] D. Sanchez and C. Kozyrakis, “Zsim: Fast and accurate microarchitectural simulation of thousand-core systems,” ACM SIGARCH Computer architecture news, vol. 41, no. 3, pp. 475–486, 2013.

[31] N. Binkert, B. Beckmann, G. Black, S. K. Reinhardt, A. Saidi, A. Basu, J. Hestness, D. R. Hower, T. Krishna, S. Sardashti et al., “The gem5 simulator,” ACM SIGARCH computer architecture news, vol. 39, no. 2, pp. 1–7, 2011.

[32] A. Hansson, N. Agarwal, A. Kolli, T. Wenisch, and A. N. Udipi, “Simulating dram controllers for future system architecture exploration,” in 2014 IEEE International Symposium on Performance Analysis of Systems and Software (ISPASS). IEEE, 2014, pp. 201–210.

[33] A. Gebregiorgis, L. Wu, C. Münch, S. Rao, M. B. Tahoori, and S. Hamdioui, “Special session: Stt-mrams: Technology, design and test,” in 2022 IEEE 40th VLSI Test Symposium (VTS). IEEE, 2022, pp. 1–10.

[34] X. Dong, C. Xu, Y. Xie, and N. P. Jouppi, "Nvsim: A circuit-level performance, energy, and area model for emerging nonvolatile memory," IEEE Transactions on Computer-Aided Design of Integrated Circuits and Systems, vol. 31, no. 7, pp. 994–1007, 2012.
[35] M. Poremba and Y. Xie, "Nvmain: An architectural-level main memory simulator for emerging non-volatile memories," in 2012 IEEE Computer Society Annual Symposium on VLSI. IEEE, 2012, pp. 392–397.
[36] M. Poremba, T. Zhang, and Y. Xie, "Nvmain 2.0: A user-friendly memory simulator to model (non-) volatile memory systems," IEEE Computer Architecture Letters, vol. 14, no. 2, pp. 140–143, 2015.
[37] J. L. NMeza, Justin and O. Mutlu., "Evaluating row buffer locality in future non-volatile main memories," in arXiv preprint arXiv:1812.06377. arXiv, 2018.
[38] S. Chung, K.-M. Rho, S.-D. Kim, H.-J. Suh, D.-J. Kim, H.-J. Kim, S.-H. Lee, J.-H. Park, H.-M. Hwang, S.-M. Hwang, J.-Y. Lee, Y.-B. An, J.-U. Yi, Y.-H. Seo, D.-H. Jung, M.-S. Lee, S.-H. Cho, J.-N. Kim, G.-J. Park, G. Jin, A. Driskill-Smith, V. Nikitin, A. Ong, X. Tang, Y. Kim, J.-S. Rho, S.-K. Park, S.-W. Chung, J.-G. Jeong, and S.-J. Hong, "Fully integrated 54nm stt-ram with the smallest bit cell dimension for high density memory pplication," in 2010 International Electron Devices Meeting, 2010, pp. 12.7.1–12.7.4.
[39] "1 gb non-volatile st-ddr4 spin-transfer torque mram." [Online]. Available: https://shorturl.at/bgsMR
[40] K. Chandrasekar, C. Weis, Y. Li, B. Akesson, N. Wehn, and K. Goossens, "Drampower: Open-source dram power & energy estimation tool," URL:http://www. drampower. info, vol. 22, 2012.
[41] A. Brokalakis, N. Tampouratzis, A. Nikitakis, I. Papaefstathiou, S. Andrianakis, D. Pau, E. Plebani, M. Paracchini, M. Marcon, I. Sourdis, P. R. Geethakumari, M. C. Palacios, M. A. Anton, and A. Szasz, "Cossim: An open-source integrated solution to address the simulator gap for systems of systems," in 2018 21st Euromicro Conference on Digital System Design (DSD), 2018, pp. 115–120.
[42] [Online]. Available: https://www.SPEC.org/cpu2017

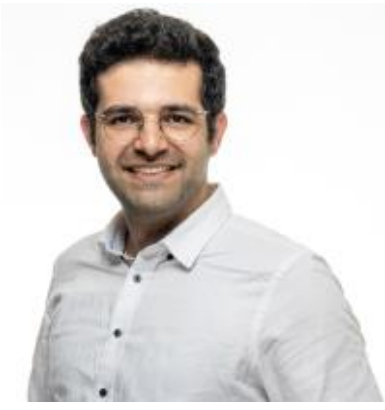

SAEED SEYEDFARAJI is a Graduate Student Member, IEEE, and holds a B.Sc. degree from Isfahan University of Technology in Isfahan, Iran, and an M.Sc. degree from Amirkabir University of Technology (Tehran Polytechnique) in Tehran, Iran. Currently, he is pursuing a Ph.D. in computer engineering at the Technische Universität Wien (TU Wien), Austria, where he also serves as a University Assistant. His research interests encompass emerging non-volatile memory technologies, in-memory processing, the integration of intelligence into hardware, and system-on-chip design. Notably, he received the Design Automation Conference 2020 Young Fellow (DAC YF 2020) Prize and was a part of the Best Team at DAC YF 2020.

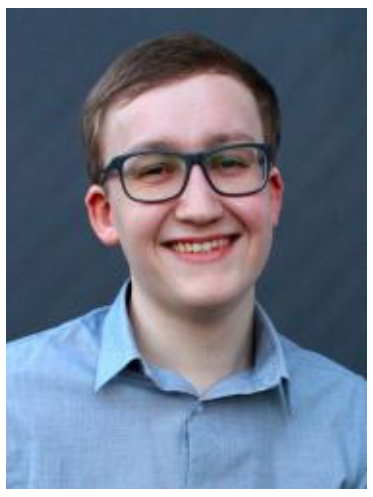

MARKUS BICHL is currently with the Technische Universität Wien (TU Wien), Faculty of Electrical Engineering and Information Technology (ETIT) as a student. He started his studies with Technische Universität Wien (TU Wien), Faculty of Informatics, Bachelore's Programme of Computer Science in 2016. He is pursuing his Master's degree in Electrical Engineering in the Master's Programme Embedded Systems. His main research interests include Emerging Memory Technologies, low-power computing, FPGA development, ASIC design, and cyber-physical systems. Besides his studies, he is working on industry-leading electrical powertrains for the automotive industry, with hundreds of thousands of units already produced. His passion is to work further on Embedded Systems topics and gain a professional career in research.

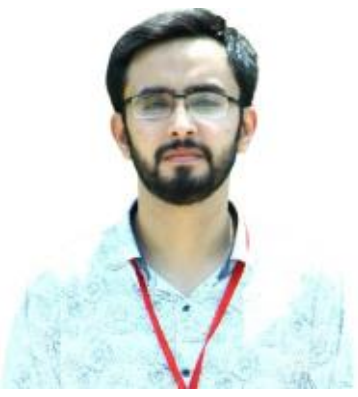

ASAD AFTAB is a Graduate Student Member, IEEE, and received the B.S. degree in Computer Systems engineering from University of Engineering and Technology (UET), Peshawar, in 2017 and the M.S. degree in Electrical (Telecommunication and Computer Networks) engineering from the National University of Sciences and Technology (NUST), Islamabad, in 2021. He is currently pursuing a Ph.D. in Electrical engineering at the Technische Universität Wien (TU Wien), Austria. His research interests encompass designing both hardware and software-based sustainable security techniques for autonomous CPS, which includes researching suitable ML algorithms for defence, analyzing various adversarial attacks, and exploring innovative defence methods to enhance the resilience of machine learning algorithms.

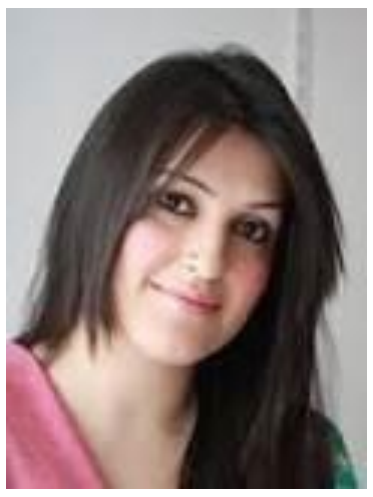

SEMEEN REHMAN is currently with the Technische Universität Wien (TU Wien), as an Assistant Professor. In October 2020, she received her Habilitation Degree in the area of Embedded Systems from the Faculty of Electrical Engineering and Information Technology, TU Wien. She has co-authored one book, multiple book chapters, and more than 60+ publications in premier journals and conferences. Her main research interests include dependable systems and energy-efficient embedded system, approximate computing, security, IoT/CPS. She has received the CODES+ISSS 2011 and 2015 Best Paper Awards, DATE 2017 Best Paper Award Nomination, HiPEAC Paper Awards, DAC Richard Newton Young Student Fellow Award, and Research Student Award at the KIT. She served as the Topic Track Chair and co-chair at the DATE and ISVLSI conferences from 2020 and 2023, and has served as the TPC of multiple premier conferences on design automation and embedded systems.